# One-dimensional ZnO exciton polaritons with negligible thermal broadening at room temperature

A. Trichet[1], L. Sun[2], G. Pavlovic[3], N.A. Gippius[3,4], G. Malpuech[3], W. Xie[2], Z. Chen[2], M. Richard[1] and Le Si Dang[1]

[1] *CEA-CNRS-UJF group Nanophysique et Semiconducteurs, Institut Néel, 25 Avenue des Martyrs, F38042 Grenoble , France*

[2] *Surface Physics Laboratory, Department of Physics, Fudan University, Shanghai 200433, China*

[3] *LASMEA, CNRS/University Blaise Pascal, 24 Avenue des Landais, 63177 Aubière, France*

[4] *A.M. Prokhorov General Physics Institute RAS, 119991 Moscow, Russia*

**Phonon damping is the main source of pure dephasing in the solid state, limiting many fundamental quantum effects to low temperature observations. Here we show how excitons in semiconductors can be totally decoupled from the phonon bath, even at room temperature, thanks to their strong interaction with photons. To do so, single ZnO microwires are investigated by angle resolved photoluminescence spectroscopy. We show that confined optical modes similar to whispering gallery modes can strongly interact with excitons to form one-dimensional (1D) quasi-particles called exciton polaritons [1-3], with normal mode splittings exceeding 300 meV. Mostly interesting is the strong quenching of the polariton-phonon interaction at room temperature, leading to a record figure-of-merit of 75 for the ratio of the Rabi splitting to the polariton full-width-at-half-maximum (FWHM). This remarkable finding, fully accountable within a**

**semi-classical modeling, opens promising prospects for the study and practical application of quantum degenerate 1D Bose gases at elevated temperatures.**

Exciton polaritons (polaritons) are bosonic quasi-particles resulting from the strong coupling between excitons – electron-hole pairs bound by Coulomb interaction in semiconductors – and electromagnetic field waves in bulk semiconductors [1] and photonic heterostructures such as Fabry-Perot microcavities [2, 3]. In recent years, microcavity polaritons have attracted increasing interest due to unique physical properties, e.g. an effective mass lighter than that of free electrons by four orders of magnitude [3]. This is the key parameter which permitted polaritons to undergo Bose-Einstein condensation at 20 K in a CdTe-based microcavity [4]. The critical temperature is actually limited by the polariton stability against high density and temperature, which directly depends on the exciton binding energy and the strength of light-matter interactions. Thus, intense effort has been carried out over the last decade to realize photonic structures using wide band-gap semiconductor materials like III-Nitrides or Zinc oxide, which offer both large exciton oscillator strength and binding energy (25 meV and 60 meV binding energy, respectively). However, the epitaxial growth of photonic structures of high structural quality is challenging with these materials, due to the lack of adapted substrates and the large lattice mismatch within the same family materials: for example, 4% of lattice mismatch between GaN and AlN and -10% between GaN and InN. Nevertheless, the strong coupling regime has been achieved at room temperature in advanced planar GaN and ZnO microcavities [5-8], and convincing indication of polariton lasing has been obtained in a Nitride-based planar microcavity [6]. In all these photonic structures, the Rabi splitting is typically 50-60 meV at room temperature, and the figure-of-merit is 4 at best, limited by both the microcavity structural quality and the large phonon damping of excitons.

Following a recent work on ZnO [9], we chose a different strategy to achieve the strong coupling regime at room temperature. Single crystalline ZnO microwires of hexagonal cross-section, with typical length of 50µm and diameter of 1µm, are grown by a vapour phase transport method under atmospheric pressure at ~ 900°C [10]. Surprisingly, this rather simple growth method provides excellent regularity of the hexagonal shape and very low surface roughness, as shown by the SEM image of the microwire used in this work in Fig.1.b. Thus high quality hexagonal-cavity whispering gallery modes (HWGM) are sustained in the microwires without any additional technological processing: quality factors of 800 are reported in this work for modes close to the excitonic transition energy. These modes are strongly coupled with the free bulk excitons lying at ~3.30 eV (A, B, and C exciton states) at room temperature [11]. In [9] a normal mode splitting as large as 300 meV was measured and confirmed in the present work (supplementary information).

To get a deeper insight on the intrinsic properties of this new kind of polaritons, we have performed space and angle resolved photoluminescence spectroscopy of single ZnO microwires at various temperatures (5-300 K). The luminescence, excited by the 325nm line of a He-Cd laser, is collected by a 0.5 NA objective. The resulting image or its Fourier transform (Fourier plane) can be projected onto the entrance slit of a monochromator, providing respectively spectrally resolved micro-photoluminescence, or a direct access to the (Momentum, Energy) dispersion of polaritons along the slit direction as well as their homogeneous linewidth for any given momentum (see method for experimental details).

Angle resolved photoluminescence at room temperature are shown in Fig.2.a and Fig2.b for two different orientations of the detection angle θ and φ, respectively (see Fig.1). Due to the translational invariance along the wire main axis z//c, the emission angle θ is related to the polariton scalar momentum by $k_z = E\sin(\theta)/\hbar c$. Thus Fig 2.a represents also the dispersion of polaritons in the (Momentum $k_z$, Energy) plane. Several well defined branches are visible, which can be separated into two families according to their linear polarization, i.e. mostly TE with respect to the main axis of the wire (electric field E perpendicular to the wire c axis, right side of the figure) or mostly TM (E//c, left side) at $k_z = 0$ ($\theta = 0$). They correspond to different lower polariton branches, which result from the coupling between A, B, and C excitons and different HWGM modes. Their dispersion features are typical of free polaritons: i), modes of high energy (i.e. closer to the exciton resonances) have smaller dispersion than lower energy modes; ii), an inflexion point shows up at $\theta \sim 40\text{-}50°$, which results from the onset of anti-crossing between the involved HWGM and the exciton level. The dispersion and polarization properties of these polariton branches are well accounted for, using a semi-classical calculation taking into account the finite momentum $k_z$ along the wire axis and the strong excitonic anisotropy in ZnO (see supplementary informations). Normal mode splittings exceeding 300 meV are deduced from this modelling, which matches that expected in the bulk material. Indeed as pointed out in [12], the overlap integral between HWGMs and excitons is close to unity in a wire, like in the bulk material.

Fig.2.b shows the polariton dispersion measured versus the angle $\phi$. Polariton modes are now found to be strictly dispersionless, i.e. strictly monomode in the plane perpendicular to the wire main axis. This is direct evidence of the 1D nature of polaritons investigated here, rarely achieved before [13] and never with such a figure-of-merit, *a fortiori* at room temperature (see below). They differ from polaritons confined along the c axis in ZnO wire cavities reported in [14, 15].

The other striking feature in Fig.2 is the sharpness of the polariton lines, with FWHM of 4 meV only for polaritons with half excitonic fraction (zero exciton-photon detuning), leading to a record figure-of-merit of 75 for the strong coupling regime at room temperature. This small linewidth seems at first glance contradictory to the phonon damping of 40 meV reported for bare excitons [16], since the polariton linewidth is usually assumed to scale as the mean value of the bare exciton and bare photon linewidths. In fact this assumption, based on the coupling between two damped classical oscillators, cannot properly account for the phonon contribution to the polariton linewidth in ZnO microwires, because the exciton-photon interaction dominates over the exciton-phonon interaction by one order of magnitude [17]. In such a case, the phonon damping should be directly evaluated in the polariton states, using the Fermi golden rule. The results, assuming polariton scatterings with a thermal bath of acoustical and optical phonons, are displayed in Figs.3a and 3b (red solid lines) for low (70 K) and high temperatures (300K), respectively. The calculations show that phonon damping dramatically increases upon increasing temperature only for polariton modes contained within the energy range [$E_X$,$E_X$-$E_{LO}$] ($E_{LO}$ ≤ 72 meV), while those at lower energy remain virtually unaffected regardless of their excitonic fraction. This behavior is due to LO-phonon scatterings of the high energy polaritons toward higher momentum free exciton states, a very efficient process owing to the very high density of states of 3D bare excitons outside the light cone. On the other hand, LO-phonon scatterings of lower energy polaritons involve only 1D polariton states, a process strongly weakened as compared to excitons, because the polariton density of states, which scales as the polariton mass, is lower by four orders of magnitude.

To further check this finding, we have investigated the temperature dependence of polariton modes in a tapered microwire (microwire 2), which allows tuning of their energies along the wire axis [9]. At low temperatures (say below 100 K), the PL spectrum consists of a polariton modes (denoted S1, …, S5 in Fig. 3a). The presence of sharp modes S3-S5 within 20 meV below the excitonic resonances indicates that the LO-phonon population is still too low. At T=300K, however, these high energy modes as well as the S2 mode part falling in the energy range [$E_X$,$E_X$-$E_{LO}$] (corresponding to positions L < 34 μm along the wire axis) are completely washed out by phonon damping (Fig. 3b). On the other hand, the S2 lower energy part of (positions L > 34 µm) remains unaffected. It can be seen in Fig. 3c that, by contrast to the excitonic resonance A but in agreement with calculations, no measureable thermal contribution to the linewidth can be observed up to room temperature for the S2 mode measured at the position L = 43 µm, in spite of the fact that its excitonic fraction $f_X$ steadily increases from 50% at 10K to 76% at 300K.

From these considerations, a criterion can be drawn for this phonon quenching to be achieved: half the normal mode splitting must exceed the LO phonon energy, i.e. $\hbar\Omega/2 > E_{LO}$. In practice, this criterion is usually difficult to meet considering the large LO phonon energy in most semiconductors. In ZnO microwires this criterion is met for the first time thanks to the very large oscillator strength combined with the microwire geometry which provides a close to unity overlap integral between excitons and photons modes.

In this report, we have shown that the strong coupling between whispering gallery modes and excitons in ZnO microwires results in the formation of 1D exciton polaritons, with record Rabi splitting of about 300 meV. We demonstrate experimentally and theoretically that these 1D polaritons can be thermodynamically decoupled from the phonon bath, a very advantageous situation to maintain high coherence at elevated temperatures. Thus, with a record exciton binding energy of 60 meV, polaritons in ZnO microwires appear as one of the most promising Bose gases for fundamental physics and practical applications. For example, by adjusting the photon-exciton detuning with the wire diameter, one can change the strength of the repulsive polariton interaction to address the various 1D physics issues, e.g. quantum fluctuations and quasi condensation [18], thermalization and quantum Newton's cradle [19], fermionization in a Tonks-Girardeau gas [20], etc... The demonstration of high quality polaritons in ZnO microwires also opens new prospects for the fabrication in the near future of ultra compact and low cost polariton "lasers" [21], ultrafast parametric amplifiers [22], or non classical source of photon pairs [23] operating at unprecedented high temperatures.

**Methods**

The ZnO microwires are spread on a glass substrate to which they remain loosely attached by electrostatic forces of Van der Walls kind. Wires of good spectral quality have been selected for the experiments. The sample is placed into a variable temperature (5-300K) cryostat of large optical aperture.

For angle resolved measurement, a long segment (~5µm length) of the microwire featuring constant diameter (like that shown Fig.3.a and 3.b for L=40µm to 45µm) needs to be found for momentum $k_z$ to be well defined. This segment is excited by the 325nm line of a CW He-Cd laser focused by an aspheric lens and passing through the rear side of the substrate. The photoluminescence is collected by a 0.5 NA NUV enhanced objective. The Fourier plane image is formed on the slit of a monochromator by a "4f" setup. For θ-resolved measurement, the wire axis is carefully set parallel to the slit by rotating the whole sample. For ϕ-resolved measurement, the wire axis is set perpendicular to the slit.

Over-heating of the wire due the optical excitation is also checked spectrally: the laser intensity is set low enough to prevent any redshift of the spectrum.

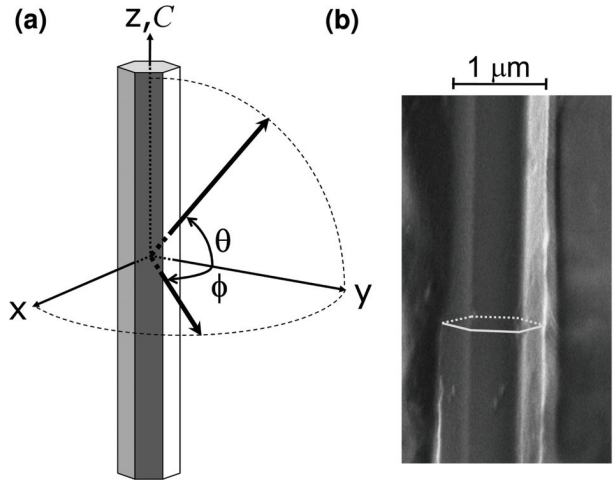

Fig. 1: a) Definition of the angles $\theta$ and $\phi$ as used in the text. b) Scanning Electron Microscope (SEM) image of the microwire under study. The measured radius is 500nm ±20nm. The gray solid and dashed line materializes the microwire hexagonal cross-section.

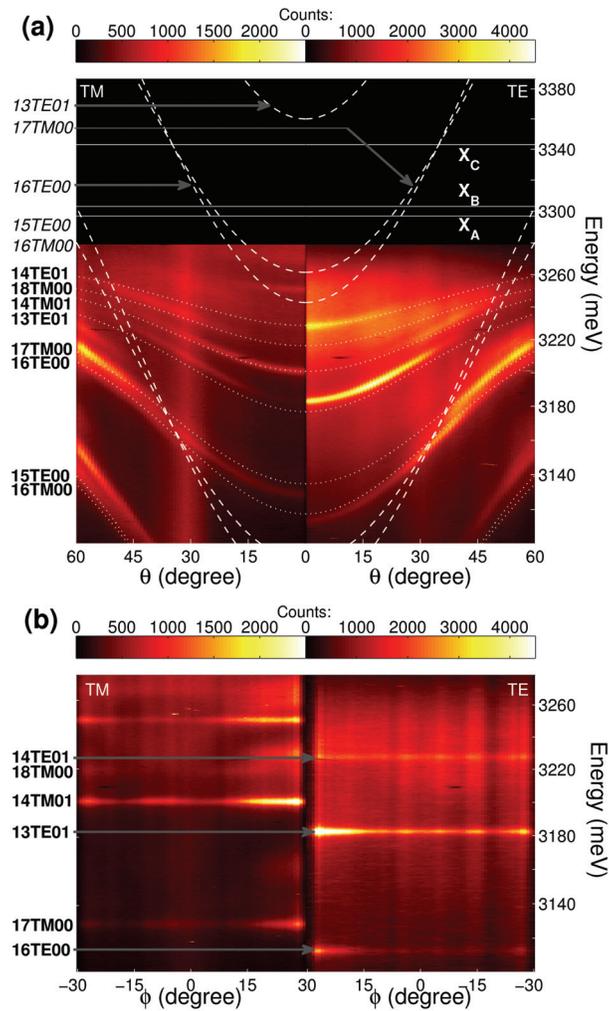

Fig. 2: Angle-resolved Room temperature photoluminescence (PL) of microwire 1 for TE (right) and TM (left) polarizations. The emission intensity is color scaled (online) and increases from black to yellow. a) PL in the ($\theta$, energy) dispersion plane. The dashed and solid lines represent the calculated dispersion of bare (uncoupled) cavity and exciton modes, respectively. The dotted lines represent the calculated polariton modes. b) PL in the ($\phi$, energy) dispersion plane. The labels on the left side of the figure refer to the polariton modes order an polarization according to the HWGM they derive from (see supplementary materials for details).

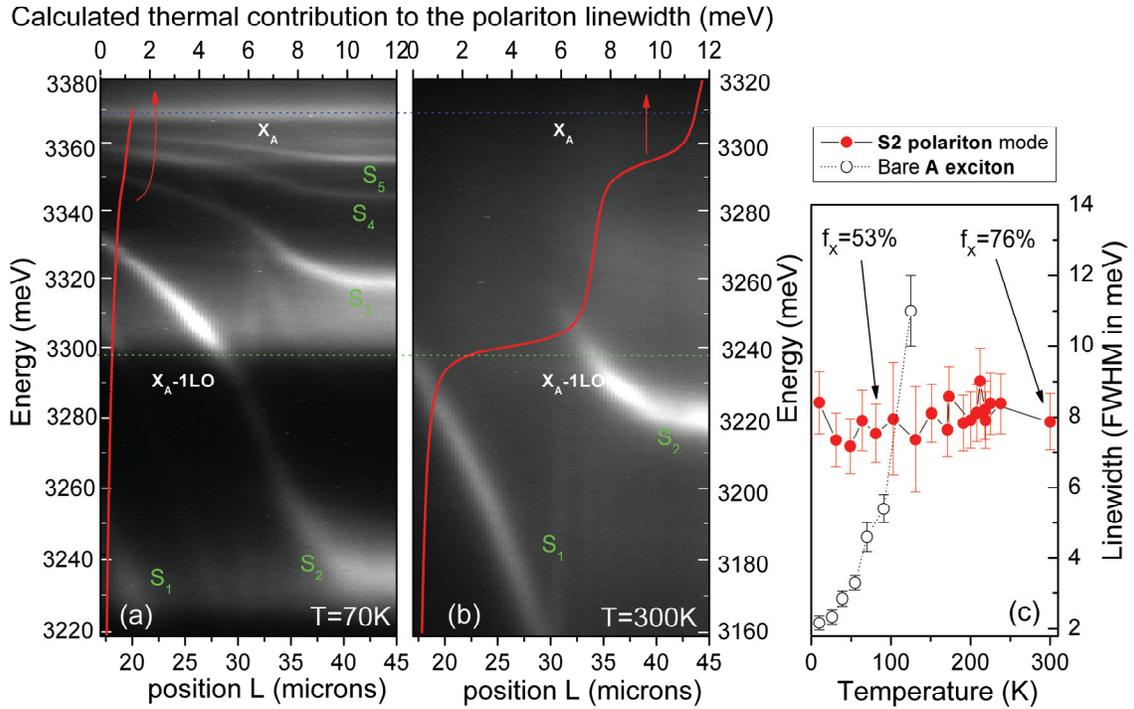

Fig. 3: Spatially resolved TE-polarized emission spectrum along 26 microns length of microwire 2 at temperature T=70K (a) and T=300K (b). Along this portion of the microwire, the inhomogeneous diameter (presently increasing from left to right) provides a natural way to continuously vary the exciton/photon detuning. $S_1$ to $S_5$ are the labels of five visible polariton

modes. The free A exciton level $X_A$ is materialized by the blue dotted line and the green dotted line figures the A exciton energy minus 1 LO-phonon energy ($E_{LO}$=72meV). The red plot shows a calculation of the phonon contribution to the polariton linewidth versus energy. At room temperature modes (b) S3, S4, S5 have vanished due to the excessive broadening taking place in the region [$X_A$, $X_A$-$E_{LO}$] while those at lower energy S1 and a part of S2 (right side of the wire) remain virtually unaffected regardless of their significant excitonic fraction. Interestingly when S2 crosses the energy boundary E= $X_A$-$E_{LO}$ at 300K (at position L=34 µm) it abruptly vanishes. (c) Plots the homogeneous linewidth of polariton mode S2 versus temperature (red filled circles) extracted from angle-resolved emission measurement carried out in a region centred on L= 43 µm. There, the S2 mode energy lies below $E_X$-$E_{LO}$. In agreement with the calculation, although its excitonic fraction $f_X$ steadily increases from 50% at 10K to 76% at 300K, no measureable thermal contribution to the linewidth builds up upon increasing the temperature to 300K. The linewidth of bare A exciton is plotted versus temperature for comparison (hollow circles).

**Supplementary Information**

**A- Exciton-polariton dispersion in cylindrical birefringent cavity**

So far the theoretical treatment of hexagonal microwires has been limited to the calculation of modes with zero momentum ($k_z$ = 0) [1-3]. It has been shown numerically [2] that modes of hexagonal microwires match those of cylindrical cavities [3] within a few percents of deviations only. Thus, to reproduce our microwire data, we solve Maxwell's equations in the cylindrical geometry, and take into account the anisotropy of the excitonic response. In cylindrical coordinates, the permittivity reads:

$$\varepsilon_{r(z)}(\omega) = \varepsilon_{\infty}\left(1 + \sum_{t=A,B,C} \frac{\omega_{t,LT}^{r(z)}}{\omega_{t,ex} - \omega - i\Gamma}\right) \quad (1)$$

where $\omega_{t,LT}^{r(z)}$ are the longitudinal-transverse splittings along the microwire axis $z$ and the radial direction $r$ in the cross-section plane, $\varepsilon_{\infty}$ and $\Gamma$ are the background dielectric constant and excitonic non-radiative decay rate, respectively, $\omega_{t,ex}$ are the A, B and C excitonic resonances. The Maxwell's equations for waves propagating along $z$ solve the transverse components in terms of the longitudinal ones [4]. They can be found from

$$\frac{\partial^2 E_z}{\partial z^2}\left(\frac{\varepsilon_z}{\varepsilon_r} - 1\right) + \Delta E_z + \frac{\omega^2}{c^2}\varepsilon_z E_z = 0, \quad (2a)$$

for TM modes, whereas the magnetic field is described by the standard Helmholtz's equation

$$\Delta H_z + \frac{\omega^2}{c^2}\varepsilon_r H_z = 0, \quad (2b)$$

for TE modes. The solutions of the equations (2a) and (2b) in the radial direction are Bessel functions of the first kind inside the microwire, and Hankel functions outside. They are the proper choices in the case of an open cavity, when the so-called "leaky" modes with small dissipative part are of interest. The $z$ and $\phi$ dependencies are purely propagative, i.e. of the form $\exp(ik_z z + im\phi)$. Each mode is determined by the propagation constant along the z-axis $k_z$, the azimuthal number $m$ and the radial number $n$, which count the number of zeros of the electric field along the wire circumference (*2m* zeros) and the radial direction (*n* zeros), respectively.

Let $a$ be the radius of the cylindrical wire of the same cross section as the hexagonal wire of radius $b$: $a = b\sqrt{3\sqrt{3}/2\pi}$. Expressing the continuity of the tangential field components at the boundary $r = a$, we can determine the constants of the problem for a fixed $n$ and $m$ and derive the dispersion relation $\omega_{n,m}(k_z)$ and the polarization degree $\rho_{n,m}(k_z)$ of each mode. Then only modes featuring the highest finesse, typically modes with *n=0* or *n=1*, are considered in the chosen energy range. For non-zero $k_z$ and $m$, there are no pure TE polarized or TM polarized modes in this geometry, even in the isotropic case. However, the calculation shows that this mixing is strongly enhanced by the excitonic anisotropy and by the strong coupling.

As shown in Fig.2.a, the main polariton characteristics of the strong coupling are well reproduced by the model, and a general agreement can be found between calculated and measured mode dispersions. The parameters used in the fit are: $a$ = 500 nm, $\varepsilon_b$=6.35, $\hbar\omega_A$=3.297eV, $\hbar\omega_B$=3.303eV, $\hbar\omega_C$=3.343eV, $\hbar\omega^r_{LT,A}$=2.7meV, $\hbar\omega^r_{LT,B}$=12.8meV, $\hbar\omega^z_{LT,C}$=16meV, and the other LT splittings are taken to be 0, i.e. A, B, and C excitons are assumed to be purely TE, TE, and TM polarized, respectively. The eight modes used in the modelling are (in order of decreasing energy) 14TE01, 18TM00, 14TM01, 13TE01, 17TM00, 16TE00, 15TE00, 16TM00 (the first number stands for $m$, the last for $n$). The dashed and solid lines in Figure 2.a are the dispersions of the bare photonic and excitonic modes, respectively. The first four modes are positively detuned, while the last four ones are negatively detuned in energy with respect to the exciton modes. Changing the microwire radius by ± 20 nm would indeed change the mode quantum numbers. However, a similar overall fit as displayed in Fig.2 can be achieved by adjusting the set of parameter values by less than ± 10%. These values are also in general agreement with those of Refs. [5-7].

The angular ($\theta$) dependence of the polarization degree of every mode in the strong coupling regime is highly non-trivial. A measurement for modes 13TE01 and 14TM01 is shown in Figures Ia and Ib. The most striking feature is the change in the polarization observed with increasing emission angle θ: the 13TE01 (14TM01) mode at θ = 0° completely switches to TM (TE) mode at θ ~ 40° (30°). This polarization switching is mainly due to the strong coupling, which mixes the cavity mode with every exciton state simultaneously. It is satisfactorily reproduced by the model (solid lines in Figs.Ia and Ib) for the TE mode but not for the TM mode. In fact, a complete agreement on this point is more difficult to achieve for two reasons: i), the polarization mixing is sensitive to the geometry of the system (our model assumes a circular and not hexagonal cavity); ii), the weak TM (TE) components of B (C) excitons have been neglected.

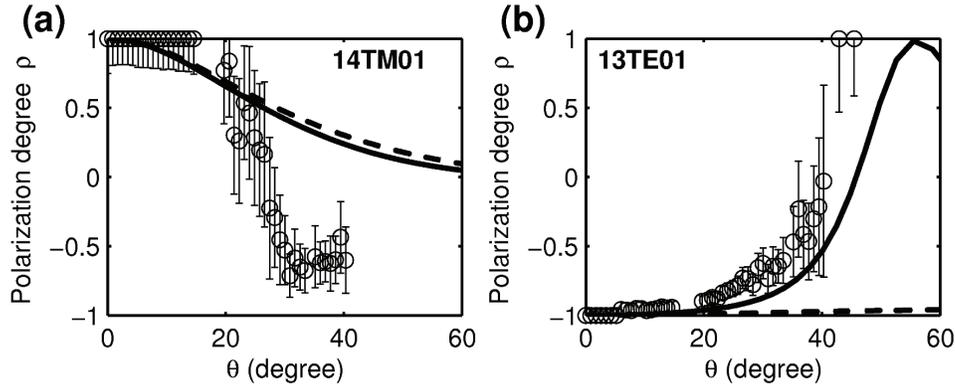

Fig. I: Angular $\theta$ dependence of the polarization degree of modes a) 14TM01 and b) 13TE01. Angular $\theta$ dependence of the full width at half maximum of modes Open symbols are the measurements, solid and dashed lines are the calculations for polariton modes and bare (uncoupled) optical modes, respectively.

B- One dimensional exciton-polaritons coupling with acoustic and LO phonons

The standard procedure used to evaluate the phonon induced broadening of an exciton-polariton line is to calculate the broadening of the bare exciton line $\gamma_X$, and then to describe the- coupling of this broadened exciton with the photon modes. In such a case, the thermal polariton broadening is given by $f_X \gamma_X$. This convenient procedure is actually quite inaccurate if $k_b T \ll \hbar\Omega/2$. The proper approach is indeed to first consider the strongest interaction in the system which is the light-matter coupling, and then to consider the phonon induced broadening of the polariton modes themselves. However, in most experimental cases studied till now, the width of the polariton modes is dominated by the radiative width and the inhomogeneous broadening. The low figure of merit normally achieved makes that the thermal contribution becomes dominant only when $k_bT$ becomes of the order of $\hbar\Omega/2$. Our sample, with its high figure of merit opens a new window in which $k_bT$ is much larger than the bare mode width and much lower than the half of the Rabi Splitting. In such a case, the description of the thermal broadening as $f_X\gamma_X$ gives the largest contribution to the mode broadening and is strongly inaccurate.

To compute the phonon contribution to the damping of a given 1D polariton state at $k_z$=0, the Fermi golden-rule is applied that includes the scattering of the polariton (with the exciton/photon detuning as a parameter) towards excitonic states (3D) by absorption or emission of acoustic and LO-phonons. The scattering toward other 1D polariton states was found negligible and is not considered. In this framework, the only parameter playing an important role is the energy difference between the initial polariton and the bare exciton states. Deformation potential for acoustic phonons and Fröhlich interaction for LO phonons [8] are considered. The deformation potential value used is D=15 eV. The LO phonon energy is 72 meV, the optical dielectric constant is 6.25 and the static dielectric constant is 8.7. A similar calculation has been done in [9] for bulk polaritons.